\documentclass[%
 aip,
 amsmath,amssymb,
 reprint,%
]{revtex4-1}

\usepackage{graphicx}
\usepackage{dcolumn}
\usepackage{bm}
\usepackage{tikz}
\usepackage{pgfplots}
\usepgfplotslibrary{colormaps}

\usepackage[utf8]{inputenc}
\usepackage[T1]{fontenc}
\usepackage{mathptmx}
\usepackage{etoolbox}

\makeatletter
\def\@email#1#2{%
 \endgroup
 \patchcmd{\titleblock@produce}
  {\frontmatter@RRAPformat}
  {\frontmatter@RRAPformat{\produce@RRAP{*#1\href{mailto:#2}{#2}}}\frontmatter@RRAPformat}
  {}{}
}%

\newcommand{\horacio}[1]{\textcolor{black}{#1}}

\makeatother

\begin{document}

\preprint{AIP/123-QED}

\title{Confinement-induced collective motion in suspensions of run-and-tumble particles}

\author{José Martín-Roca}%
\altaffiliation{These authors equally contributed to this work}
\affiliation{%
 Universidad Complutense de Madrid\\
 Departamento de Estructura de la Materia, F\'isica T\'ermica y Electr\'onica
}
\affiliation{%
 \horacio{GISC – Grupo Interdisciplinar de Sistemas Complejos, 28040 Madrid, Spain}
}

\author{Daniel Escobar Ortiz}
 \altaffiliation{These authors equally contributed to this work}
 \affiliation{%
 Universidad Complutense de Madrid\\
 Departamento de Estructura de la Materia, F\'isica T\'ermica y Electr\'onica
}

\author{Chantal Valeriani\textsuperscript{$*$}}
\affiliation{%
 Universidad Complutense de Madrid\\
 Departamento de Estructura de la Materia, F\'isica T\'ermica y Electr\'onica
}
\affiliation{%
 \horacio{GISC – Grupo Interdisciplinar de Sistemas Complejos, 28040 Madrid, Spain}
}

\author{Horacio Serna\textsuperscript{$*$}}%
\affiliation{%
 Universidad Complutense de Madrid\\
 Departamento de Estructura de la Materia, F\'isica T\'ermica y Electr\'onica
}
\affiliation{%
 \horacio{GISC – Grupo Interdisciplinar de Sistemas Complejos, 28040 Madrid, Spain}
}
\email{hoserna@ucm.es, cvaleriani@ucm.es}

\date{\today}
             
\begin{abstract}
Collective motion is ubiquitous in active systems at all length and time scales. The mechanisms behind such collective motion usually are alignment interactions between active particles, effective alignment after collisions between agents or symmetry-breaking fluctuations induced by passive species in active suspensions. In this article, we introduce a new type of collective motion in the shape of a traveling band induced purely by confinement, where no explicit or effective alignment are prescribed among active agents. We study a suspension of run-and-tumble particles confined in microchannels comprising asymmetric boundaries: one flat wall and one array of funnel-like obstacles. We study the phase behavior of the confined active suspension upon changes in the packing fraction and the persistence length to define the stability region of the traveling band. We characterize the traveling band structurally and dynamically and study its stability with respect to the geometry of the microchannel. Lastly, we describe the mechanism of motion of the band, which resembles the tracked locomotion of some heavy vehicles like tractors, finding that a counter-flux of active particles in the lower part of the band, explained in terms of source-sink and vacancy diffusion mechanisms, is the facilitator of the traveling band and sustains its motion. We name this new collective phenomenon \textit{confinement-induced tracked locomotion}.  
\end{abstract}

\maketitle


\section{\label{sec:introudction}Introduction}

Collective motion is ubiquitous in Nature at different length and time scales \cite{vicsek2012collective}, ranging from genetic material migration during cell division \cite{scholey2003cell}, to bacterial swarming \cite{kearns2010field,darnton2010dynamics}, the mesmerizing collective migration 
of
\textit{Dictyostelium}\cite{hashimura2019collective},  the phototactic response of the green micro-algae \textit{C. reinhardtii} \cite{witman1993chlamydomonas,arrieta2017phototaxis,choudhary2019reentrant,choudhary2025transient}, 
swarms of insects \cite{topaz2008model,ariel2015locust}, schools of fish \cite{niwa1994self,lopez2012behavioural,liu2025collective}, flocks of birds \cite{bialek2012statistical,cavagna2014bird}, herds of sheep \cite{garcimartin2015flow}, and crowds of people \cite{gu2025emergence}, among other examples.  

Thirty years ago,  Vicsek and collaborators proposed 
a model to mimic the collective behavior of aligning active particles: 
a simple aligning mechanism between point-like agents moving in a noisy environment, affecting the agents' orientation
\cite{vicsek1995novel}. In particular, at intermediate densities and noise intensities, the Vicsek model predicted the formation of traveling bands (dense regions of aligning particles)\cite{chate2008collective,chate2008modeling,solon2015pattern,ginelli2016physics}. The alignment between particles was the underlying mechanism of such traveling bands. The effects of confinement on the collective motion of active particles with alignment interactions have recently been addressed. Aligning self-propelled particles confined in arrays of obstacles with different shapes and distributions has revealed trapping and sub-diffusive behavior \cite{chepizhko2013diffusion,chepizhko2015active}. On the other hand, confinement in ordered arrays of obstacles can induce a re-entrant collective behavior, leading from self-organization into traveling bands to a disordered state and, as the packing fraction of obstacles increases,  to a new synchronous collective behavior \cite{serna2025influence}.   

Although the Vicsek model can qualitatively explain many collective phenomena observed in organisms with evolved sensory organs and complex signal processing systems, it cannot do the same for their microscopic counterparts. 
This is due to colloids or motile microorganisms usually exhibiting collective motion as a response to light or chemical stimuli \cite{budrene1995dynamics,wu2015collective,erban2004individual,liebchen2018synthetic,keller1971traveling, arrieta2017phototaxis, choudhary2019reentrant,choudhary2025transient} or due to hydrodynamic interactions \cite{alarcon2013spontaneous,alarcon2017morphology,zottl2014hydrodynamics}, rather than through an explicit alignment mechanism among agents. Thus, systems exhibiting collective motion that arises without aligning interactions between agents and without the action of external fields are highly appreciated in the active matter community and provide a platform for understanding the physics behind this emergence.

Some examples of systems with no explicit aligning interactions that exhibit collective motion in  bulk are "energy-depot" active disks which  form traveling bands as a consequence of an effective alignment mechanism mediated by collisions \cite{miranda2025collective}, and active-passive mixtures of Brownian particles that form propagating interfaces as a result of a complex combined mechanism involving source/sink effects and oriented displacements of  particles modulated by density fluctuations induced by the presence of passive particles \cite{fernandez2025dynamics,wysocki2016propagating}.

Geometrical confinement has  proven to be an efficient mechanism in rectifying \cite{anand2024transport,tailleur2009sedimentation,wan2008rectification,di2010bacterial}, trapping and sorting \cite{galajda2007wall,martinez2020trapping,kaiser2012capture,kumar2019trapping} active particles, especially configurations comprising concave regions in which the active particles tend to accumulate, such as funnels or chevrons \cite{reichhardt2017ratchet}. Moreover, the shape of the confining boundaries may induce jamming states\cite{pajger2025jamming} or destroy phase separation\cite{ben2022disordered} in active systems. In this article, we present a simulation study of a two-dimensional suspension of run-and-tumble particles confined in asymmetric microchannels with one flat wall and one wall made of funnel-like obstacles (see Fig. \ref{fig:scheme}).

To the best of our knowledge, we report for the first time the emergence of a dense traveling band whose formation and motion are completely induced and sustained by confinement, with neither explicit alignment between agents, nor effective alignment induced by particle shape or collisions, nor by the presence of passive particles in the suspension. We discovered such a traveling structure serendipitously, while performing the analysis of the results published in Ref.\cite{serna2025sorting} by some of the present authors. The mechanism of the band's motion resembles the tracked locomotion of some heavy vehicles such as tractors, and thus we have named it \textit{confinement-induced tracked locomotion}.

This article is organized in the following way. In section \ref{sec:model}, we introduce the model and the analysis tools. Section \ref{sec:results} contains the results and the discussion, and Section \ref{s:conclusions} includes a summary and the conclusions. Finally, Section \ref{s:SI} has an index for the \textbf{Supplementary Information}.  

\section{\label{sec:model}Model and methods}

\subsection{Model}

\begin{figure}[h!]
    \centering
    \includegraphics[width=0.35\textwidth]{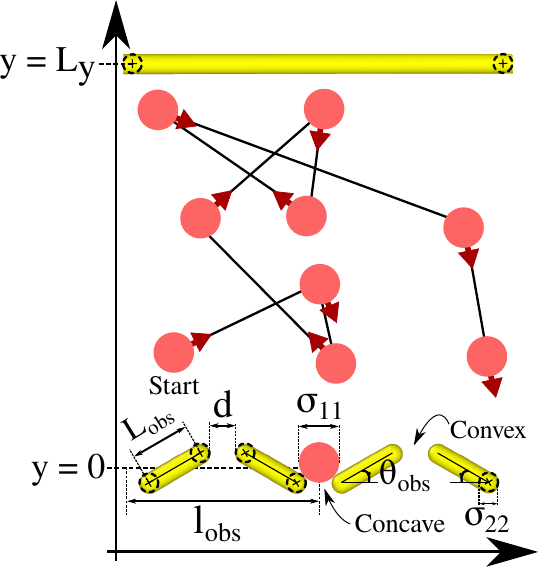}
    \caption{Scheme of the geometry of the system. The top boundary of the microchannel is a flat wall, and the bottom boundary of an array of funnel-like obstacles with period $l_{obs} = 2\left( L_{obs}\cos\theta_{obs} + d + \sigma_{22}\right)$, where $d = 1.25\sigma_{22}$ is the gap size, $L_{obs} = 3.0\sigma_{22}$ is the length of the obstacles and $\theta_{obs}$ the angle formed between the obstacles and the $x$-axis. The walls of the microchannel are composed of frozen particles (in yellow) of diameter $\sigma_{22}$, with a lattice constant, $l_w = 0.12\sigma_{22}$. The axis $y = 0$ is placed at the center of the obstacles. The active particles (in pink) follow the run-and-tumble phenotype observed in different motile microorganisms.}
    \label{fig:scheme}
\end{figure}

The system under consideration is a two-dimensional suspension of active particles (type 1, in pink in Fig. \ref{fig:scheme}) following run-and-tumble dynamics confined in microchannels composed of frozen particles (type 2, in yellow in Fig. \ref{fig:scheme}). \horacio{Figure \ref{fig:scheme} shows a scheme of the system with a graphic explanation of the geometrical parameters of the channel.} 

Run-and-tumble motion is implemented following Langevin dynamics:
\begin{equation}
    m_i\frac{d^2 \textbf{r}_i}{dt^2} = F_a\textbf{n}_i -  \boldsymbol{\nabla}_i U -\gamma\,\textbf{v}_i + \sqrt{2\gamma m_i k_BT}\,\textbf{W}_i\, .
    \label{e:langevin}
\end{equation}

Here, $\boldsymbol{r}_i$, $\boldsymbol{n}_i$ and $\boldsymbol{v}_i$ are the position vector, unit orientation vector and velocity vector of particle $i$, respectively. The active self-propulsion force is set to $F_a = 10$, the friction coefficient, $\gamma = 10$, the mass of the particles $m_i = 1.0$, and the thermal energy, $k_BT = 1.0$. The self-propulsion speed of the particles is, thus $v_0 = F_a/\gamma = 1.0$. With such a choice of parameters, the inertia of the particles' motion is only significant at very short time scales, so we can consider the system follows overdamped dynamics \cite{serna2025sorting}.  The thermal bath applies a stochastic force, $\boldsymbol{W}_i = (W^x_i,W^y_i)$, on the $i$th particle, given by the standard Gaussian white noise with zero mean and correlation $\langle W_i^x(t)W^y_i(t)\rangle = \delta_{ij}\delta_{xy}\delta(t - t')$. 

All particles interact  via a WCA\cite{weeks1971role} repulsive potential,
\begin{equation}
    U(r) =
    \begin{cases} 
       4\epsilon_{mn}\left[ \left ( \frac{\sigma_{mn}}{r} \right)^{12} -\left (\frac{\sigma_{mn}}{r} \right)^6 \right] + \epsilon_{mn}, & r< 2^{1/6}\sigma_{mn}, \\
        0 & r \geq 2^{1/6}\sigma_{mn}.
    \end{cases}
    \label{e:WCA}
\end{equation}
with indexes $m$ and $n$ running over the two particle types \horacio{(1: active particles, 2: wall particles)}. We set $\sigma_{11} = 2$, $\sigma_{22} = 1$ and $\epsilon_{11} = \epsilon_{22} = 1$. The crossed interactions are given by the Lorentz-Berthelot mixing rules\cite{lorentz1881ueber,berthelot1898melange}: $\sigma_{mn} = (\sigma_{mm} + \sigma_{nn})/2$ and $\epsilon_{mn} = \sqrt{\epsilon_{mm}\epsilon_{nn}}$. Particles constituting the microchannel walls are placed at a distance $l_w = 0.12 \sigma_{22}$ from each other.  \horacio{Note that wall particles (type 2) do not interact with each other, they only interact with active particles (type 1).}

Run-and-tumble motion is simulated as a discrete stochastic process, where  the tumbling events are described by:
\begin{equation}
\theta_i(t + dt) = 
    \begin{cases} 
       \theta_i(t), & \alpha < \zeta {(t)} \\
        \theta_i(t) + \Phi{(t)}  & \alpha > \zeta {(t)}
    \end{cases} ,
    \label{e:run-and-tumble}
\end{equation}
being $\theta_i$  the angle formed between the orientation vector $\textbf{n}_i$ and the positive $x$-axis. $\alpha = t_{trial}/\tau_r$ is the probability of occurrence of a tumbling event,  $t_{trial}$ being the time between trials, and $\tau_r$  the mean time between tumbling events or reorientation time. Here we use $t_{trial} = dt$, i.e. tumbling trials are performed at every step. Thus, the tumbling rate is $\alpha_r = \alpha/t_{trial} = 1/\tau_r$. $\zeta \in [0,1]$ is a uniformly distributed random number generated at each step and $\Phi \in [0,2\pi]$ is a uniformly distributed random angle. 

The particles' level of activity  is thus characterized by the persistence length, $l_p = \frac{F_a}{\alpha_r\gamma}$, which is the average distance an active particle travels before tumbling. In our work, we vary the suspension's packing fraction, $\phi = \frac{N\pi\sigma_{11}^2}{4L_xL_y}\in [0.20,0.40]$, with $N$ the number of active particles and $L_x$ and $L_y$ the edge lengths of the rectangular simulation box. We fix \horacio{$L_y = 10\sigma_{11}$ for most of the simulations. We will state it explicitly when chosen  otherwise}. The system contains $1000 \leq N \leq 20000$ active particles, depending on packing fraction and system size. $L_x$ is chosen and adjusted accordingly. The initial condition is defined as a random distribution of active particles inside the microchannel with a uniform random distribution of orientations. Periodic boundary conditions are imposed on the $x$-axis. The confining channel comprises funnel-like obstacles in the lower part and a flat wall in the upper part (see Fig. \ref{fig:scheme}). The obstacles feature gaps that do not allow the passage of the active particles. The gaps provide two stable resting positions for the active particles, since they can fit in the gaps (but cannot go through them). 
In Fig. \ref{fig:scheme}, we present the geometry of the funnel-like obstacles: we set their length to $L_{obs} = 3.0\sigma_{22}$, forming an angle $\theta_{obs}$ with respect to the $x$-axis which will be varied in the range $[0^o, 75^o]$. The width of the gaps is fixed to $d = 1.25\sigma_{22}$. The period of the obstacle array is given by $l_{obs} = 2\left( L_{obs}\cos\theta_{obs} + d + \sigma_{22}\right)$.
For comparison, we also consider \horacio{three} different channel types: \horacio{the first} with two flat parallel walls, \horacio{the second} consisting of a top flat wall and a bottom wall made of corrugations with no gaps (see \textbf{Fig. S1} of \textbf{Supplementary Information}) \horacio{, and the third composed of a top wall made of trapezoids and a bottom wall comprising the original funnel-like obstacles with gaps (see Fig.\ref{fig:Vcm-thetaObs}b)} . 

All quantities in the manuscript are expressed in reduced simulation units, using $k_BT$, $\sigma_{22}$, and $m$ as units of energy, distance and mass, respectively. The simulations are performed using the open-source Molecular Dynamics package LAMMPS\cite{plimpton1995fast,LAMMPS}, in which we have implemented the run-and-tumble dynamics in equation \ref{e:run-and-tumble}. The equations of motion are integrated using the NVE integrator and the Langevin's thermostat with $dt = 0.001\tau$, where $\tau = \sqrt{\frac{m\sigma_{22}^2}{k_BT}}$ is the simulation time unit. Most of the simulations are run for up to $30000\tau$, but some are run for longer to check the stability of some of the observed self-organized steady states.

\subsection{Analysis tools}\label{sec:Analysis Tools}

\subsubsection{Identifying dense structures: Percolation}

To start with, we characterize the internal structure of the system and, by analyzing the instantaneous particle configurations, we identify structures such as traveling bands and clogs: the former consist of a large percolating cluster moving along the channel, while the latter consists of a large percolating immobile cluster. For each configuration, the set of pairwise particle distances was evaluated under periodic boundary conditions along the longitudinal direction of the channel (x-axis) and fixed boundaries along the y-axis.   Two particles were considered connected if their separation was smaller than a prescribed cutoff distance $r_\mathrm{cut}= \sigma_{11}$, chosen slightly above the mean interparticle spacing within the dense phase as shown in the radial distribution function (see  \textbf{Fig. S2, Supplementary Information}). From this connectivity criterion, a percolation network was constructed, where each particle represents a node and pairwise connections define edges. Clusters were then identified as the connected components of this network, and the system is considered in a percolated state when the network connects particles from one extreme, $y=0$, to the other, $y=L_y-\sigma_{12}$, along the y-axis.  
Having established the largest connected cluster in the system, this will correspond to the traveling band or the clog case for the following analysis. This procedure provides a robust identification of the collective structure. It allows the calculation of different observables, such as the temporal evolution of the band size and the distribution of orientation angles within the dense region.

\subsubsection{Following the structures: Band's velocity}

The propagation speed of the traveling band was determined from the evolution of its center of mass (CM) along the channel axis. Given that the system is periodic in the longitudinal direction, the CM position was computed using an angular (circular-mean) formulation that prevents discontinuities when the band crosses the periodic boundaries. Each particle position $x_i$ was mapped to a phase angle $\theta_i = 2\pi(x_i - x_{\min}) / L_x$, from which the mean direction in angular space was calculated as  
\[
\tan\theta_\mathrm{CM} = \frac{\langle \sin\theta_i \rangle}{\langle \cos\theta_i \rangle}.
\]
The corresponding CM coordinate was then obtained by inverting this transformation back to real space. The coordinate $x_\mathrm{CM}(t)$ thus obtained provides a smooth measure of the band’s translational motion, from which the instantaneous and mean propagation velocities were extracted. As shown later, the CM velocity seems to be well defined with a constant value close to $\langle V \rangle \approx \frac{1}{3} \, v_0$. We will use this value to characterize the motion's mechanism.

\begin{figure*}[ht!]
    \centering
    \includegraphics[width=\linewidth]{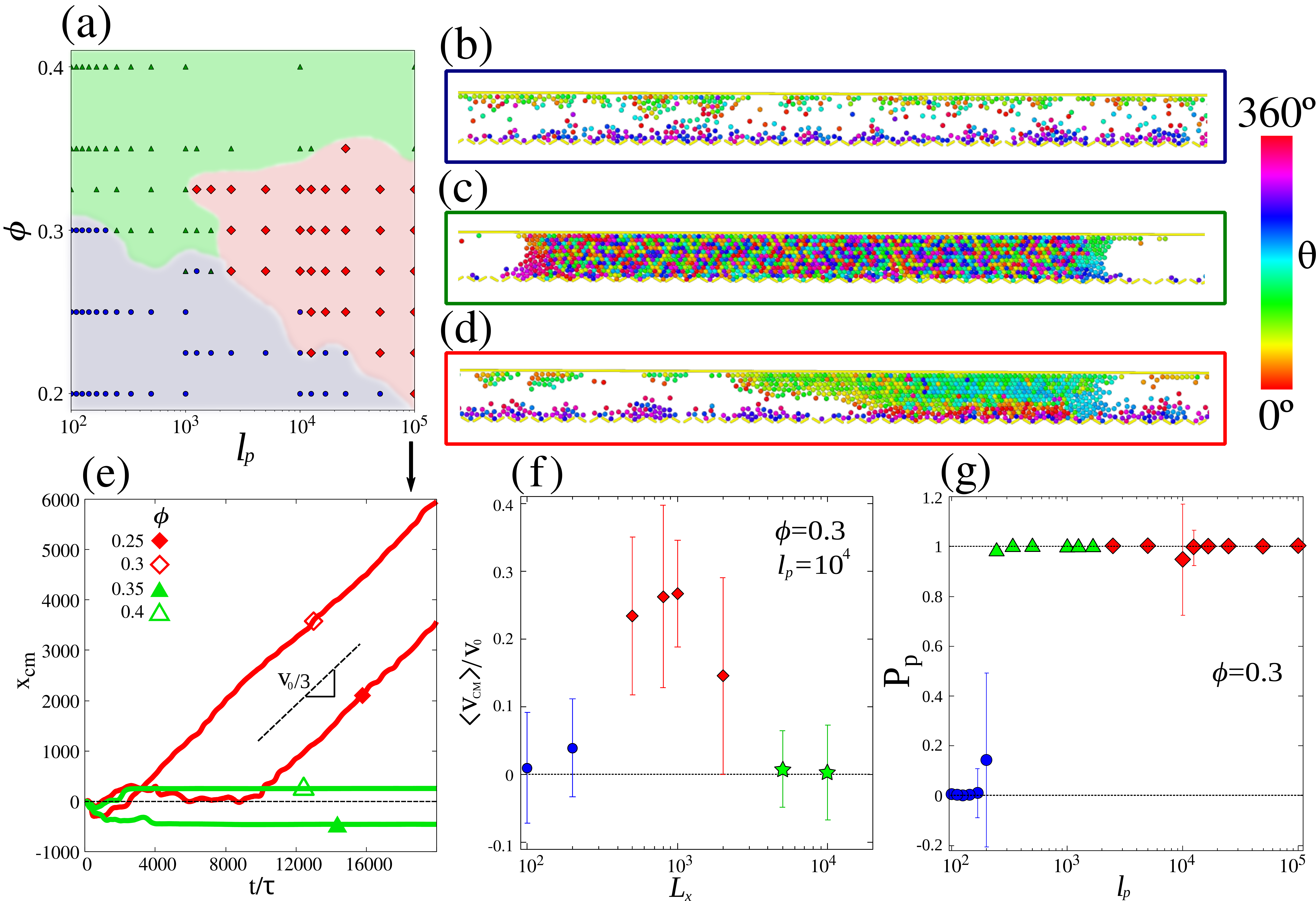}
    \caption{Phase behavior of the confined active suspension.\textbf{(a)} Dynamic phase diagram in the plane $\phi-l_p$ for the active suspension confined in a microchannel with funnel-like obstacles configured at $\theta_{obs}=30^\circ$ \horacio{and $L_x = 797\sigma_{22}$}. The different phases are represented by symbols and colors.\horacio{ The areas corresponding to each phase have been colored as a rough estimate to provide a guide for the eye.} \textbf{(b)} Non-percolated system (blue circles). \textbf{(c)} Clogging ( green triangles). \textbf{(d)} Traveling band (red diamonds). The color code of the active particles in \textbf{b}-\textbf{d} represents the  active particle's orientation angle, $\theta$, with respect to the $x$-axis, according to the color bar on the right. \horacio{\textbf{(e)} Evolution of the $x$ component of the center of mass of the system, $x_{cm}$, at  $l_p=10^5$ and different packing fractions, $\phi$. \textbf{(f)} Average center of mass velocity of the system, $\langle V_{CM} \rangle$, at persistent length $l_p=10^4$ and packing fraction $\phi=0.3$ for different channel lengths. Green stars markers used for long channels represent systems where more than two traveling bands form, very often moving in opposite directions. At long times, these traveling bands merge to form clogs. \textbf{(g)} Average percolation probability, $P_p$, as a function of $l_p$ at $\phi=0.3$. The color of the points correspond to the phases presented in panel \textbf{a}.  }  }
    \label{fig:diagrams}
\end{figure*}

\subsubsection{Motion mechanism: Fluxes and internal particle displacements in the traveling band}\label{sec:fluxes}

Finally, to quantify the microscopic dynamics within the traveling band, we evaluated the local flux and density profiles, as in Ref.\cite{fernandez2025dynamics}. We employ a coarse-grained formulation based on Smoluchowski’s equation for the instantaneous density and current fields in two dimensions.   The microscopic density is defined as $\rho(\mathbf{r},t) = \left\langle \sum_i \delta(\mathbf{r} - \mathbf{r}_i(t)) \right\rangle$ and currents are defined as $\mathbf{j}(\mathbf{r},t) = \left\langle \sum_i \delta(\mathbf{r} - \mathbf{r}_i(t))\, \mathbf{v}_i(t) \right\rangle$, where $\langle \cdots \rangle$ denotes an ensemble average.   These fields satisfy the Smoluchowski continuity equation, $    \frac{\partial \rho(\mathbf{r},t)}{\partial t} + \nabla \cdot \mathbf{j}(\mathbf{r},t) = 0$. During the steady propagation of the traveling band, the center of mass of the dense phase moves with an average constant velocity $\langle V \rangle \approx \frac{1}{3} \, v_0$ along the longitudinal direction of the channel.   Assuming that the band profile is self-averaging in the comoving frame, the coarse-grained density can be written as $\rho(\mathbf{r},t) \simeq \tilde{\rho}(\tilde{\mathbf{r}})$, where $\tilde{\mathbf{r}} = \mathbf{r} - \langle \mathbf{V} \rangle t=( x - \langle V\rangle t,y)$ and $\tilde{\rho}(\tilde{\mathbf{r}})$ is the stationary density distribution in the reference frame of the traveling band. In this frame, the continuity equation reduces to a balance between the advective and diffusive fluxes, $\nabla_{\tilde{\mathbf{r}}} \cdot \mathbf{j}(\tilde{\mathbf{r}}) 
=  \langle \mathbf{V} \rangle \cdot \nabla_{\tilde{\mathbf{r}}} \tilde{\rho}(\tilde{\mathbf{r}}),$ which can be formally integrated to obtain the local current density as
\begin{equation}
    \mathbf{j}(\tilde{\mathbf{r}}) 
= 
\langle \mathbf{V} \rangle\, \tilde{\rho}(\tilde{\mathbf{r}}) - \boldsymbol{\Delta}(\tilde{\mathbf{r}}),
\end{equation}
where $\boldsymbol{\Delta}(\tilde{\mathbf{r}})$ represents a counter-current term that compensates the advective transport produced by the collective motion of the band, and must satisfy $\nabla_{\tilde{\mathbf{r}}} \cdot \boldsymbol{\Delta}(\tilde{\mathbf{r}})=0$. Due to the fact that the system is traveling mainly along the $x$-direction of the channel, we can focus only on the longitudinal component of the previous equation,  averaging out its dependence in $x$ to explore the fluxes of the band along the perpendicular axis 
\begin{equation}
    {j}_x(y) = \langle V \rangle\; {\rho}(y) - {\Delta}_x (y)
    \label{e:finalfluxes}
\end{equation}

In contrast with previous cases, where the counter-current is a constant \cite{fernandez2025dynamics}, we observe a dependency with $y$ due to the presence of the funnel-like obstacles in the bottom boundary of the channel. This is an educated assumption based on direct observations and preliminary analysis of the traveling band trajectories. The counter-current accounts for the internal rearrangements and local diffusive fluxes of particles within the dense region, which reduce the apparent current below the purely advective value $\langle \mathbf{V} \rangle\, \tilde{\rho}$.  
Therefore, the spatial dependence of $\boldsymbol{\Delta}(\tilde{\mathbf{r}})$ provides direct information on the internal dynamics and flux balance inside the traveling band.

To measure the flux, successive configurations separated by a fixed time interval $\Delta t=1 \, \tau $ were used to estimate particle velocities from finite differences of their positions.  Particles were binned along the transverse ($y$) direction into strips of thickness $\delta  y =\sigma_{11}$, and for each bin we computed the local density $\rho(y)$ and the longitudinal flux component $j_x(y)$.   The counter-current, defined as the difference between the advective and self-propulsive contributions to the flux, ${\Delta}_x (y) = \langle V \rangle\; {\rho}(y) - {j}_x(y)$, was used to characterize the degree of internal rearrangement within the band.   Averaging these quantities over time yields stationary profiles that provide detailed information on the internal mass transport and collective motion in the traveling structure.

\section{Results\label{sec:results}}

\subsection{Phase behavior\label{sec:phase behaviour}}

We start by exploring the phase behavior  of the active suspension confined in microchannels with $\theta_{obs} = 30^o$ in the $l_p-\phi$  plane \horacio{($l_p = \frac{F_a}{\alpha_r\gamma}$, $\phi = \frac{N\pi\sigma_{11}^2}{4L_xL_y}$)} , as presented in Figure \ref{fig:diagrams}a, where  the different phases are: non-percolating clustering state  (blue circles, panel b),  clogging (green triangles, panel c) and traveling band (red diamonds, panel d). \horacio{To construct the phase diagram, we consider channels with $L_x = 797\sigma_{22}$}. Videos of these states are available in the \textbf{Supplementary Video 1}. \horacio{All the graphics of trajectories of simulations presented in this article were generated in part using the visualization software OVITO \cite{stukowski2009visualization}.}

To classify, from a structural perspective, the different self-organized steady states observed in the active suspension, we determine the probability of finding a percolating configuration along the $y$-axis at the steady state, $P_p$. High values of $P_p$ indicate states where a big dense cluster spans the system in the $y$-direction, whereas low values correspond to states in which the active particles are not forming such structures. \horacio{In Figure \ref{fig:diagrams}g, we present $P_p$ as a function of $l_p$ for $\phi = 0.30$, illustrating the different self-organized states with the same color code as in the phase diagram (see Fig. \ref{fig:diagrams}a) .} The dense structures \horacio{($P_p\approx 1.0$)} may correspond to clogs or traveling bands. To distinguish between these two, we monitor the speed of the center of mass of the whole system in the x direction. \horacio{Figure \ref{fig:diagrams}e shows the temporal evolution of the $x$ component of the position of the center of mass, $x_{cm}$}. States corresponding to a clog are characterized by \horacio{ constant $x_{cm}$ and thus by} velocities  close to zero, in contrast to states corresponding to traveling bands \horacio{, in which $x_{cm}$ exhibits well-defined slopes and thus velocities different than zero}. Combining the percolation and the center of mass velocity criteria, we build the phase diagram in Fig. \ref{fig:diagrams}a. \horacio{Note that the traveling states presented in Fig. \ref{fig:diagrams}e (in red), correspond to bands moving to the right. However, we observe that the traveling bands move indistinctly to the right and left, as the microchannel geometry lacks features that would bias their motion. We will discuss this in more detail later.}

The clustering state consists of active particles accumulating at the funnels or at the flat wall of the microchannel and takes place at low $l_p$ and $\phi$ (see Fig. \ref{fig:diagrams}b). \horacio{The clog state occurs for suspensions with high $\phi$ (see Fig. \ref{fig:diagrams}c). As $l_p$ increases, the formation of the traveling structure is favored, especially for intermediate packing fractions ($\phi \leq 0.325$). The clogging state} also appears  when  confining the suspension  in channels with two flat walls (with the same  $l_p$ and $\phi$), and can be acknowledged as the confined equivalent of motility-induced phase separation (MIPS) in a bulk suspension with periodic boundary conditions \cite{cates2015motility,knippenberg2024motility} (see \textbf{Figure S1a}, \textbf{Figure S1b} and \textbf{Supplementary Video 2}). The traveling band is observed in persistent active suspensions ($l_p>4\times10^3$) and intermediate to low packing fractions ($0.30\leq\phi\leq0.20$), and consists of a big cluster of active particles that moves persistently towards left or right. In the example  shown in Fig. \ref{fig:diagrams}d, it moves to the left.   

Interestingly, the traveling band is only stable under confinement in microchannels featuring either funnel-like obstacles with gaps or corrugations. Neither in bulk nor in confinement within channels with two flat walls the band is stable, suggesting that its formation is induced by the asymmetry of the obstacles in one of the confining walls (see \textbf{Figure S1} and \textbf{Supplementary Video 2}). We also observe that, at low packing fractions, ($\phi \leq 0.25$), the system exhibits a clustering state for low $l_p$ that directly becomes a traveling band as $l_p$ increases. Contrastingly, at higher packing fractions, ($\phi\geq 0.275$), the system shows clogging behavior before the traveling band is formed as $l_p$ increases (see Fig \ref{fig:diagrams}a).

\horacio{Finally, we study the stability of the traveling band as a function of the channel's length, $L_x$. In Figure \ref{fig:diagrams}f, we illustrate the average speed of the center of mass of the system, $\langle V_{CM} \rangle$, upon increasing $L_x$, for $\phi = 0.30$ and $l_p = 10^4$, conditions at which the traveling band is stable as shown in Fig. \ref{fig:diagrams}a. For short channels ($L_x \lessapprox 500\sigma_{22}$), the traveling structure does not fully develop. In the range $500\sigma_{22}\lessapprox L_x \lessapprox2000\sigma_{22}$, one traveling structure fully develops and exhibits persistent motion for long time scales. For $L_x \gtrapprox 5000\sigma_{22}$ we observe $\langle V_{CM} \rangle$ close to zero. However, these states do not strictly correspond to clogs, since many traveling structures form at the beginning of the simulations, moving, very often, in opposite directions. As time evolves, these traveling structures eventually merge into clogs. And thus $V_{CM}$ approaches zero. We mark these states in Fig.\ref{fig:diagrams}f with green, the characteristic color of clogging states in the phase diagram of Fig. \ref{fig:diagrams}a, but with stars instead of triangles, indicating that at the beginning of the simulations the traveling structures are formed with a mean lifetime of a few thousands of $\tau$. Although these traveling bands observed in long channels survived for shorter time scales, their mechanism of motion is exactly the same as the one exhibited by the bands formed in shorter channels with $500\sigma_{22}\lessapprox L_x \lessapprox2000\sigma_{22}$ in which only one stable band is formed. Since the objective of this article is to characterize the mechanism of this novel collective motion induced by confinement, we keep $L_x$ within the range in which only one stable traveling band is observed, and thus have, in turn, enough statistics for the analysis.}

\subsection{Characterization of the traveling band}\label{sec:band cahracterization}

Having explored the phase behavior of the system, we now focus on the features of the traveling band. First, we calculate the speed of the system's center of mass in the $x$ direction, $V_{CM}$, for cases in which a clog or a traveling band is formed. Note that the center of mass of the system roughly matches that of the largest cluster when the system is in a clogging or a traveling state. 

\begin{figure}[h!]
    \centering
    \includegraphics[width=0.45\textwidth]{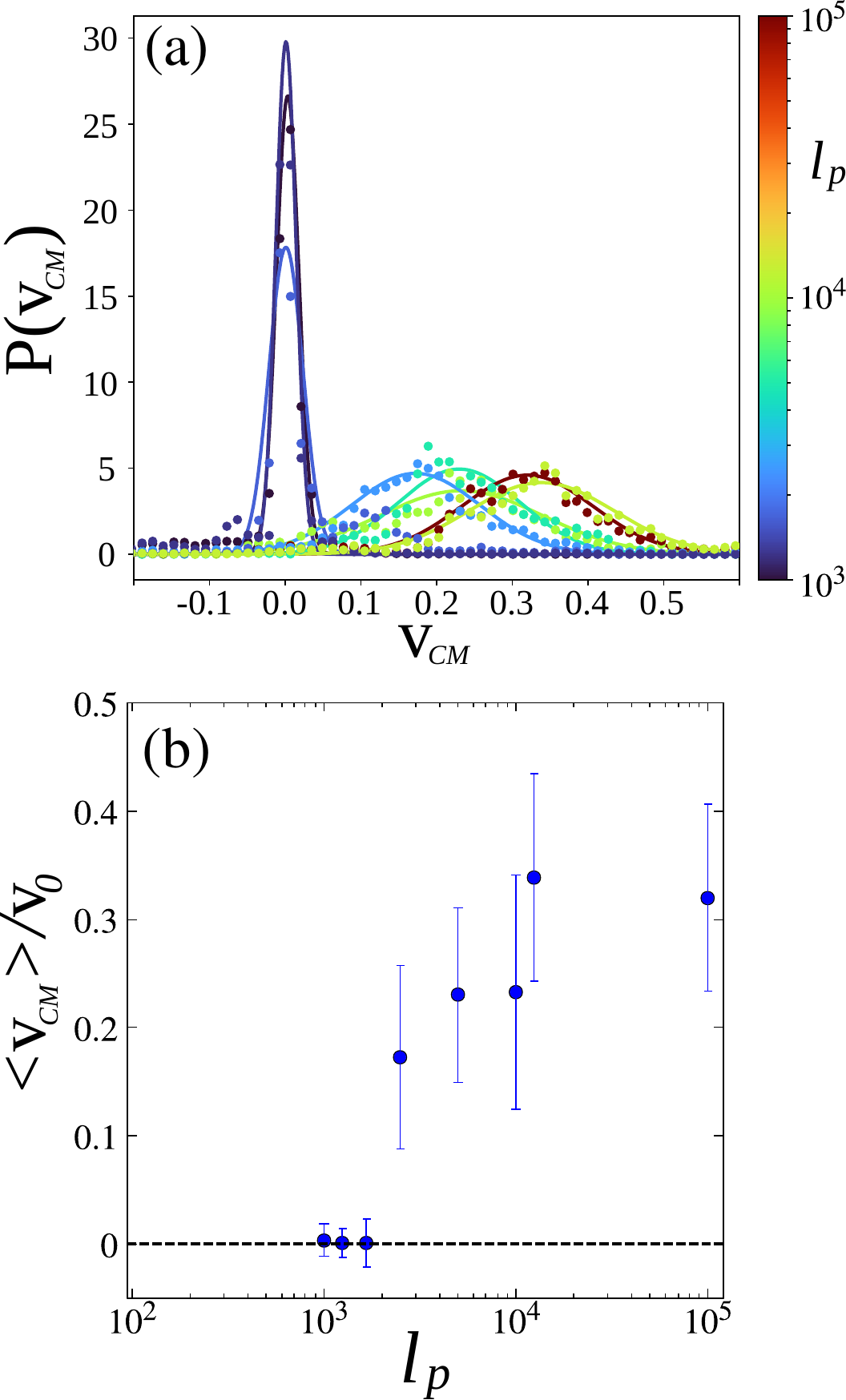}
    \caption{ Center of mass velocity of the system \textbf{(a)} Probability distributions of the Center-of-mass velocity of the system for $\phi=0.3$ and $\theta_{obs} = 30^{\circ}$, at different values of $l_p$, for clogging and traveling states. \textbf{(b)} Average Center-of-mass velocity of the system as a function of the persistent length, $l_p$.}
    \label{fig:VeloCM}
\end{figure}

In Fig. \ref{fig:VeloCM}a, we present the probability distributions of $V_{CM}$. The distributions calculated from simulations (circles) fit well to Gaussian distributions (solid lines). The color bar represents the persistence length, $l_p$. For clogging states, $P(V_{CM})$ is centered at $0$ with a small standard deviation, whereas, for traveling states, the distributions are shifted to non-zero values with higher standard deviations. Clearly, the mean speed of the center of mass, $\langle V_{CM}\rangle$, can be used as an order parameter to monitor the transition from clogging to traveling states, as presented in Fig. \ref{fig:VeloCM}b. Remarkably, the traveling band reaches average speeds, $\langle V_{CM}\rangle$, of up to $35\%$ as high as the active particles' self-propulsion speed (see Fig. \ref{fig:VeloCM}b). \horacio{It should be noted that this motion in one direction has been studied without obstacles in systems with active-passive mixtures \cite{fernandez2025dynamics} or in the presence of activity with an asymmetric spatial dependence in the system \cite{metzger2024revisiting, metzger2025exceptions}, such that a traveling band is observed. In the latter cases, a preferred direction emerges in the system, leading to a ratchet effect on the motion.}

\horacio{In contrast to the literature, the system studied here does not exhibit left-right asymmetry, and the appearance of a net motion was not obviously expected. The motion is not induced by a ratchet effect, since at the limit of $t\to \infty$ we would expect zero flux, as shown by direction reversals that compensate for long enough time scales (see \textbf{Fig S4}). However, we stress that the observed traveling states can be very persistent, with a time life of hundreds of thousands of simulation time units, as shown in \textbf{Fig. S4}. If we consider that the moving band typically has a velocity of $\langle V_{CM} \rangle \approx v_0/3=\sigma_{11}/(6 \, \tau)$ and can persist in a given direction for $\sim 150,000\, \tau$ (see \textbf{Fig S4}), then the entire band can travel $\sim 22,500$ body lengths of active particles, which, in terms of any microorganism performing a run-and-tumble motion, represents an enormous collective persistence.}

\begin{figure*}[ht!]
    \centering
    \includegraphics[width=16cm]{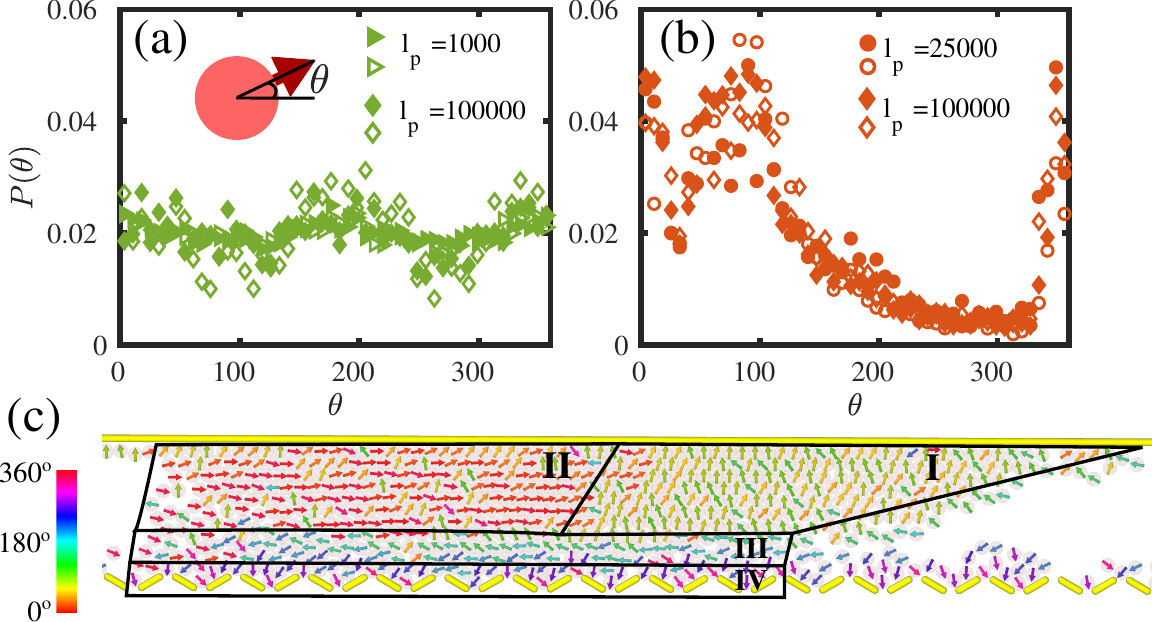}
    \caption{ Probability distributions of the orientation angle , $\theta$, of active particles for four selected cases. \horacio{The scheme of the active particle shows the definition of $\theta$ as the angle between the orientation vector of the active particle with respect to the x-axis.} \textbf{(a)} Clogging states. Filled markers correspond to $\phi = 0.40$, and empty to $\phi = 0.35$. \textbf{(b)} Traveling states. Filled markers correspond to $\phi = 0.275$ and empty to $\phi = 0.25$. \textbf{(c)} Orientation domains within the band. \textbf{I}: Particles pointing mainly upwards \horacio{, acting as a break of the traveling band}. \textbf{II}: Particles pointing towards the direction  of motion of the traveling band, constituting the thrust. \textbf{III}: Particles forming a thin layer in the lower part of the band pointing opposite to the band's motion. This domain is the facilitator of the collective motion. \textbf{IV}: Particles trapped in the funnels mainly pointing downwards. The distributions are computed using only the orientations of the largest cluster in the system. The band moves to the right and $\theta_{obs} = 30^{\circ}$.}
    \label{fig:P_theta}
\end{figure*}

With the goal of unraveling the mechanism behind the traveling band's motion at the macroscopic level, we calculate and compare the orientation distributions, $P(\theta)$, of the active particles in clogging and traveling states (see Fig. \ref{fig:P_theta}). 

In Fig. \ref{fig:P_theta}a we report $P(\theta)$ for clogging states at different $l_p$ and $\phi$, observing, as could be expected, that the distributions are rather uniform. This is already visible from the snapshot presented in Fig. \ref{fig:diagrams}c, where, at the clog boundaries, active particles point in opposite directions, while inside the clog they are oriented at random. The profile of the clog boundaries seems parabolic, which could be due to the effective attraction towards the walls induced by activity. On the other hand, the $P(\theta)$ for traveling states in Fig. \ref{fig:P_theta}b, also at different $l_p$ and $\phi$, exhibits two distinctive peaks around $\theta = 90^o$ and $\theta = 180^o$. We have chosen cases in which the traveling structure moves to the right as shown in Fig. \ref{fig:P_theta}c, although the simulations show no clear preference for any direction. As seen in Fig. \ref{fig:P_theta}b, despite the different values of $l_p$ and $\phi$, as long as those values are within the stability region of the traveling band, the distribution of orientations inside the band remains the same, indicating the same underlying mechanism.

Figure \ref{fig:P_theta}c shows a typical snapshot of the traveling band, with the orientation vectors of the particles colored according to the angle they form with respect to the x-axis. We  identify four domains  corresponding to the features of the $P(\theta)$ shown in Fig. \ref{fig:P_theta}b. Domain \textbf{I} is composed of particles mainly pointing upwards and roughly containing  half of the band's particles, thus corresponding to the peak around $\theta = 90^o$. Domain \textbf{II} is also composed of approximately half of the band's particles; the particles in this domain point in the direction of motion of the band (the peak around $\theta = 0^0$ and $360^o$ due to periodicity of $\theta$) and thus represent the propulsion force of the band. In other words, domain \textbf{II} pushes domain \textbf{I}.  
Domain \textbf{III} is composed of the particles below domains \textbf{I} and \textbf{II}, that form a thin layer that spans the whole band along its $x-$axis and points in the opposite direction of the band, corresponding to a short peak around $\theta = 180^o$. Finally, domain \textbf{IV} is mainly composed of particles trapped in the concave region of the gaps which are pointing downwards. Particles that belong to domain \textbf{IV} are reflected in the $P(\theta)$ as the flat region in the range $225^o<\theta<325^o$.

\subsection{Mechanism of traveling band's motion\label{sec:mechanism}}

The well-defined internal structure consisting of four  domains is maintained as long as the band moves. Any drastic modification of this structure leads to the formation of a clog (high $\phi$) or a clustering state (low $\phi$), depending on the packing fraction. \horacio{}

Although it is clear that the band's motion relies on domain \textbf{II} pushing domain \textbf{I}, the dynamics of domains \textbf{III} and \textbf{IV} are also crucial for understanding the mechanism that sustains this motion. As the band moves, there is an inflow of particles directed opposite to the band's motion. As shown in Fig. \ref{fig:P_theta}c, the profile of the boundary of the traveling band, which faces the above-mentioned inflow of particles, can be described as a straight line whose slope has the same sign as the band's direction of motion. Thus, the inflow of particles can move down along the band's linear boundary to reach its lowest part, filling the gap between the top domains (\textbf{I} and \textbf{II}) and the particles trapped in the obstacles. The band keeps moving on top of the particles that constitute the inflow (now domain \textbf{III}), which, in turn, are arranged on top of the particles trapped in the gaps (now domain \textbf{IV}). Particles in domain \textbf{III} find their way out, hopping through vacancies opened due to the relocation of particles induced by the motion of domains \textbf{I} and \textbf{II}, and the sliding along the obstacles. Finally, particles of domain \textbf{III} leave the band from behind. Following the described mechanism, the same particles populate \textbf{I} and \textbf{II} for extended periods, with low exchange rate, whereas domains \textbf{III} and \textbf{IV} are always composed of different particles. In the \textbf{Supplementary Video 3}, we present a movie with the details of the band's motion.

As \horacio{domain \textbf{II}} is the thrust of the band, the counter-flow of particles represented in domain \textbf{III} is the facilitator of its motion. If  particles in domain \textbf{III} could not flow through the lower part of the band, \horacio{usually because of an internal rearrangement of the band's structure,} it would lead to the accumulation of particles opposing to the band's motion in its frontal part, \horacio{balancing out the thrust force exerted by domain \textbf{II} and, in consequence,} ending up in a clogging state. \horacio{ A detailed study on the dynamics of the traveling-clogging transition is left for future work.} Particles in domain \textbf{III} are able to find their way out of the band due to the presence of the obstacles that comprise two stable positions for active particles with a height difference (convex and concave gaps), which facilitate hopping and sliding. This is confirmed by simulations of confined active suspensions inside channels with two flat walls, where only clogging states form (see \textbf{Figure S1}).

\begin{figure*}[ht!]
    \centering
    \includegraphics[width=\linewidth]{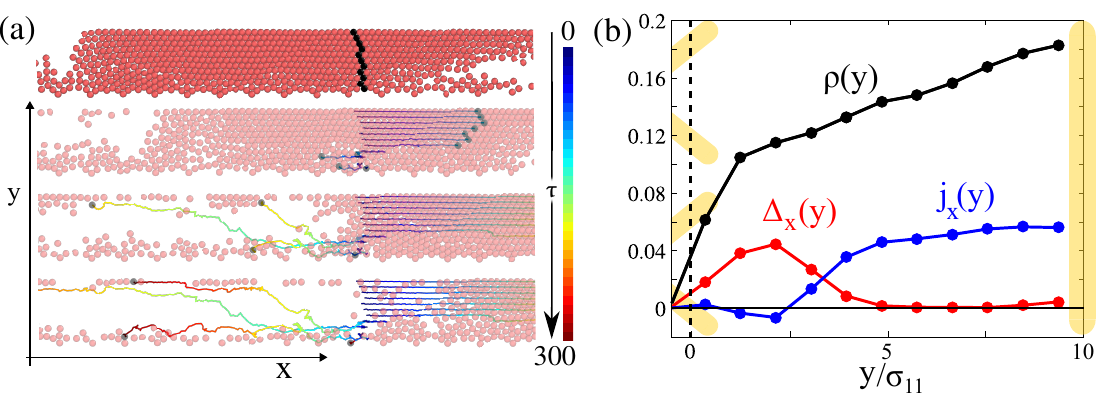}
    \caption{ \textbf{(a)} Representative particle trajectories depending on their initial positions along the $y$-axis inside the traveling band, for $l_p = 10^5$, $\phi = 0.3$ and $\theta_{obs} = 30^\circ$. Wall particles are not shown for the sake of clarity. \textbf{(b)} Corresponding local density profiles $\rho(y)$, total flux $j_x(y)$, and counter-current \horacio{$\Delta_x(y)$} along the $y$-axis. The channel walls are depicted in yellow for reference.}
    \label{fig:Fluxes}
\end{figure*}

Since the motion of the bands implies two currents of particles moving in opposite directions- domains \textbf{I} and \textbf{II} in the band's direction and \textbf{III} in the opposite direction - it somehow resembles the tracked locomotion performed by some vehicles, although both currents move at different speeds (as will be discussed later in Fig. \ref{fig:Fluxes}). Thus, we name this new type of collective motion \textit{confinement-induced tracked locomotion}. 

Active systems are well known to exhibit strong finite size effects, especially those forming traveling structures \cite{kursten2020dry,chate2020dry,fernandez2025dynamics}. With the intention of testing the robustness of our results, we perform simulations setting the conditions we expect a traveling band, but for channels with $ l_p< L_x < 10l_p$. We observe that the traveling bands are still formed and move following the same mechanism described above. However, the traveling bands are stable in a shorter time scale and thus more difficult to analyze (see \textbf{Fig. S3}). \horacio{The reason behind this reduced stability is the formation of many traveling bands within the long microchannel, some of them moving in opposite directions, that eventually collide to form clogs}. 

\begin{figure*}[ht!]
    \centering
    \includegraphics[width=0.98\textwidth]{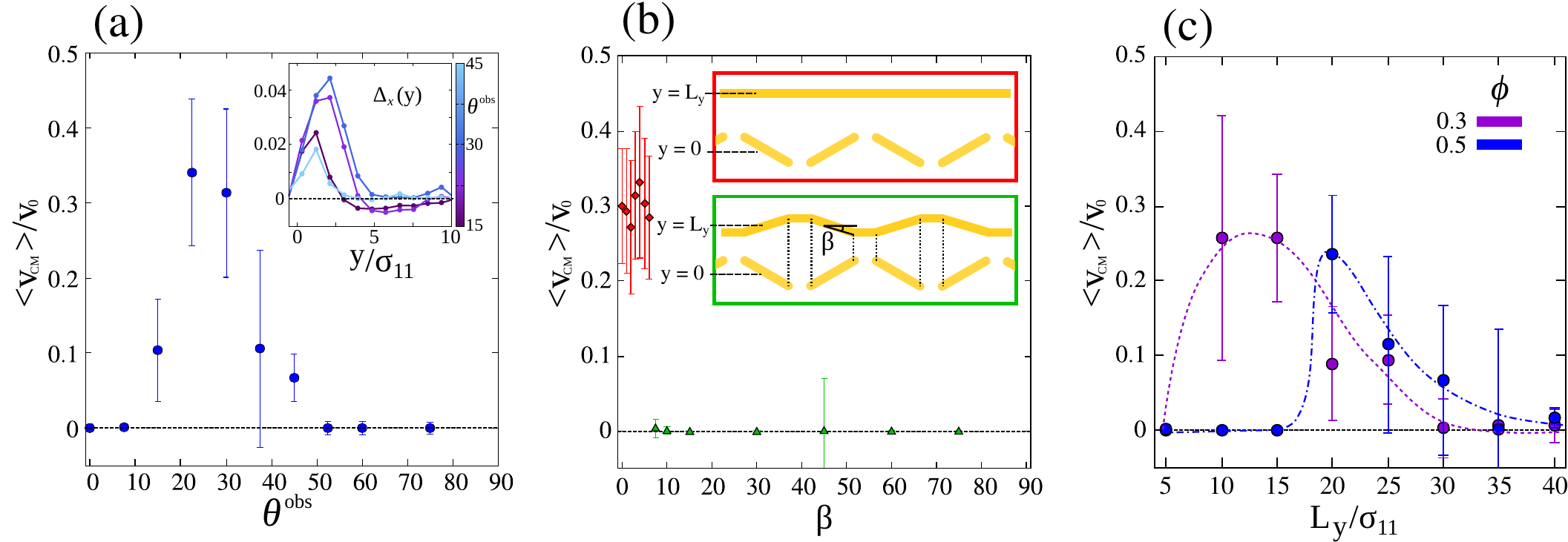}
    \caption{Center-of-mass velocity as a function of: \textbf{(a)} Corrugation tilt angle when $L_y = 10\sigma_{11}$ and $\phi = 0.30$. Inset represents the dependence of the x-component of the counter-current along the y-axis, $\Delta_x(y)$, on tilt angle, when the traveling band is formed; \textbf{(b)} top wall's corrugation angle, $\beta$ \horacio{(\textbf{see Supplementary Video 4})}; and \textbf{(c)} the width of the channel, $L_y$, when $\theta_{obs} = 30^o$ for $\phi = 0.30$ and $0.50$. The lines are a guide for the eyes to correctly observe trends. Note that the angle $\beta$ defines a gradual transformation of the top wall from flat to a trapezoidal wave which is half a period out of phase with respect to the funnel-like obstacles in the bottom wall, thus the maxima of the top wall coincide with the minima of the bottom wall and vice versa.}
    \label{fig:Vcm-thetaObs}
\end{figure*}

Aiming to characterize, at the quantitative level, the mechanism of \textit{confinement-induced tracked locomotion}, we calculate the internal fluxes of particles within the traveling band with respect to its center of mass as described in Section \ref{sec:fluxes}, following the integrated version of the Smoluchowski's continuity equation presented in equation \ref{e:finalfluxes}. 
In Figure \ref{fig:Fluxes}a, we present the trajectories of particles belonging to the band placed along the $y$-axis. We observe that particles in the top part of the band, those belonging to domains \textbf{I} and \textbf{II}, follow approximately horizontal trajectories in the direction of the band's motion, and are kept inside the band. On the other hand,  particles in the lower part of the band (domains \textbf{III} and \textbf{IV}) also describe horizontal trajectories, in the opposite direction of the band's motion, but at a lower speed. Their trajectories exhibit the characteristic pattern of vacancy diffusion. Such vacancies appear due to the rearrangement of particles caused by the motion of the upper part of the band. Particles also take advantage of sliding along the obstacles to occupy the vacancies, as shown in some of the trajectories depicted in Fig. \ref{fig:Fluxes}a. Finally, the majority of  particles in domain \textbf{III} leave the band, whereas some of them get trapped in the gaps.

In Figure \ref{fig:Fluxes}b, we present the three averaged quantities involved in equation \ref{e:finalfluxes}, for a typical system exhibiting \textit{confinement-induced tracked locomotion}: local density profile, $\rho(y)$ (in black); flux of particles in the $x$ direction along the $y$-axis, $j_x(y)$ (in blue); and the counter-current of particles, \horacio{$\Delta_x(y)$} (in red). The funnel-like obstacles and the top wall of the channel are depicted in the figure in yellow, for reference. In the concave regions of the obstacles, $\rho(y)$ adopts values close to zero. This is because only one particle per $l_{obs}$ is counted due to the shape of the obstacles. Then, $\rho(y)$ grows fast, up to positions just above the convex gaps of the obstacles (spanning domains \textbf{IV} and \textbf{II}), and then it  grows linearly, with a smaller slope, up to the top wall of the microchannel (domains \textbf{I} and \textbf{II}). As expected, $j_x(y)$ shows a small but negative value in the region that corresponds to the flow of particles belonging to domain \textbf{III} in the opposite direction of the band's motion, whereas the region comprising domains \textbf{I} and \textbf{II} ($y \geq 3.75$) shows a constant high flux of particles in the same direction of the band's motion. Both the tendencies of $\rho(y)$ and $j_x(y)$ are consistent with the structure and dynamics of the band as shown in Fig. \ref{fig:Fluxes}a (see also \textbf{Supplementary Video 3}).

Finally, $\Delta_x(y)$ shows a peak centered at the region corresponding to domain \textbf{III}, whereas it vanishes in the rest of the microchannel. As shown in a previous work \cite{fernandez2025dynamics}, a non-zero counter-current term, $\Delta_x(y)$, signals a source-sink effect that, in the case of \textit{confinement-induced tracked locomotion}, occurs in the lower part of the band and reflects the dynamics of domain \textbf{III}: particles are captured at the front end of the band and expelled from the back end of the band. Additionally, $\Delta_x(y)$ contains all the contributions that oppose to the advective transport represented by the collective motion of the band ($\langle V_{CM}\rangle \neq 0$), thus accounting for the vacancy diffusion of particles due to rearrangements within the band and facilitated by sliding along the obstacles. The vacancy diffusion mechanism that contributes to the motion of particles in domain \textbf{III} is justified since the particles inside the band are arranged in a hexagonal lattice that is continuously distorted, generating empty lattice spaces that are subsequently occupied by other particles.

\subsection{Stability of the band with respect to \horacio{ confining geometry}}

With the goal of understanding the role of \horacio{confinement} in \horacio{the emergence of traveling states, we now focus on varying the geometrical parameters of the confining microchannels. In particular, we study the stability of the traveling structure when changing $\theta_{obs}$ while keeping the top wall flat (see Fig. \ref{fig:Vcm-thetaObs}a); changing the angle $\beta$ that controls the shape of trapezoids in the top wall as shown in the inset of Fig. \ref{fig:Vcm-thetaObs}b while $\theta_{obs} = 30^o$ is maintained constant; and changing the width of the channel $L_y$ whereas $\theta_{obs} = 30^o$ is kept constant and the top wall kept flat.} 

\horacio{Let us start discussing the influence of $\theta_{obs}$.} We perform simulations at $\phi = 0.30$ and $l_p = 10^5$ for $\theta_{obs} \in [0^o,75^o]$. In Fig. \ref{fig:Vcm-thetaObs}a, using $\langle V_{CM}\rangle$ as an order parameter, we observe the existence of a range $15^o\leq\theta_{obs}\leq 45^o$ which allows the formation of traveling structures. Interestingly, for $22.5^{\circ}\leq \theta_{obs} \leq \ 30^{\circ}$ the traveling band is fully formed and reaches its top speed. As we highlighted before, the counter-current of active particles in domain \textbf{III} is the facilitator of the the collective motion, and thus, when we monitor $\Delta_x(y)$ as $\theta_{obs}$ varies, we observe a non-monotonic behavior that perfectly correlates with the behavior of $\langle V_{CM} \rangle$, exhibiting the highest peaks, precisely, for $\theta_{obs} = 22.5^{\circ}$ and $30^{\circ}$. Thus, the larger the counter-current the faster the traveling band, which might seem counter-intuitive in the first place. However, since $\Delta_x(y)$ accounts for rearrangement and diffusive fluxes opposing to the band's motion, we suggest that those angles favor particles' rearrangements at the lower part of the band by enhancing sliding and thus producing more vacancies for particles in domain \textbf{III} to diffuse into. \horacio{We also consider channels in which  the period of the obstacles at the bottom wall, $l_{obs}$, is increased. To do so, we need to vary both $L_{obs}$ and $\theta_{obs}$ simultaneously to keep the amplitude of the obstacles, $A = L_{obs}\sin(\theta_{obs} + \sigma_{22})/2$, constant. We take the case used to built the phase diagram in Fig. \ref{fig:diagrams}a as reference, with $\theta_{obs} = 30^o$, $L_{obs} = 3\sigma_{22}$ and $d = 1.25\sigma_{22}$, for which $l^{ref}_{obs} = 9.70\sigma_{22}$. We study cases with $l_{obs} = nl^{ref}_{obs}$, $n$ being a positive integer, and packing fraction, $\phi = 0.30$. We observe that increasing $l_{obs}$ in general favors clogging. At $l_{obs} = 2l_{obs}^{ref}$, a traveling band appears for $2300\tau$ to finally form a clog. Note that as $l_{obs}$ increases, the geometry of the channel approaches the limit of two flat walls that, as already discussed, only promotes the formation of clogging states. Examples of these systems can be found in \textbf{Supplementary Video 5}}

\horacio{Next, we consider a special case in which the top wall is gradually deformed by increasing the angle $\beta$ (see Fig. \ref{fig:Vcm-thetaObs}b), which converts the top wall into a trapezoidal wave, while keeping $\theta_{obs} = 30^o$. \horacio{In the inset of Fig. \ref{fig:Vcm-thetaObs}b, in the upper panel, we present the geometry of the channel when $\beta = 0^o$, whereas, in the bottom panel, we present the geometry of the channel for an arbitrary $\beta > 0^o$}. As for the funnel-like obstacles in the bottom wall for which $y = 0$ is placed to cross the mid points of the obstacles, $y = L_y$ is placed to cross the mid points of the trapezoids. Fig. \ref{fig:Vcm-thetaObs}b shows that \textit{confinement-induced tracked locomotion} only takes place for very low $\beta$ angles. For $\beta\geq7.5^o$, the active particles self-organize into a big clog. On the one hand, very small values of $\beta$ resembles the case of a flat top wall and thus recovers the traveling structure, remarking the importance of asymmetric walls to stabilize it. On the other hand, as $\beta$ increases, the active particles cannot slide freely along the top wall anymore and experience a higher effective friction coefficient when move close to the top wall due to the presence of the trapezoids, which promotes the formation of clogs. Regarding the stability of the traveling bands, now we can say that the sliding along the top walls of particles belonging to domains \textbf{I} and \textbf{II} is fundamental to sustain the \textit{confinement-induced tracked locomotion}.} \horacio{The microchannels in the inset are inside colored rectangles following the same color code as in the phase diagram of Fig. \ref{fig:diagrams}a (red for traveling states and green fro clogging states), and thus indicating that channels with small $\beta$ favor the formation of the traveling band, whereas higher values favor the formation of clogs.}

\horacio{Finally, we study the influence of the channels width, $L_y$ on the formation of traveling bands. We perform simulations of systems with $\theta_{obs} = 30^o$ and $L_x = 797\sigma_{22}$ while varying $L_y\in[5\sigma_{11},40\sigma_{11}]$. We report $\langle V_{CM}\rangle$ as a function of $L_y$ in Fig. \ref{fig:Vcm-thetaObs}c. First, we note that very narrow channels only allow the formation of clogs, which is reflected in $\langle V_{CM} \rangle \approx 0$. When $\phi = 0.30$, we observed fully-developed \textit{confinement-induced tracked locomotion} for $10\sigma_{11}\leq L_y \leq 15\sigma_{11}$ , whereas for $20\sigma_{11}\leq L_y \leq25\sigma_{11}$ a big cluster is formed and slides along the top wall. Even though the cluster is large, its size is not enough to span the channel in the $y$ direction and thus forming a traveling band. Such a cluster reaches speeds around $10\%$ of the self-propulsion speed of individual particles, and can be understood as a precursor of the traveling band only containing domains \textbf{I} and \textbf{II} (see Fig.\ref{fig:P_theta}c). For $L_y \geq 30\sigma_{11}$ a clustering state with no persistence, as that one described in the phase diagram of Fig. \ref{fig:diagrams}a is observed.  We have learned   that the size of the cluster containing domains \textbf{I} and \textbf{II} is important to the formation of traveling states within the microchannel. In order to get more insights on this matter, we repeat the previous analysis but now for $\phi = 0.5$ to induce the formation of larger clusters. As expected, at $\phi = 0.50$ those states that correspond to a persistent cluster sliding on the top wall at $\phi = 0.30$ become fully-developed traveling structures, as can be seen, for example, for $L_y = 20\sigma_{11}$. Again, as $L_y$ increases, the fully-developed traveling state converts into a persistent cluster to finally become a non-persistent clustering state. Although these are still preliminary results, we can state, to certain extent, that increasing $L_y$ shifts the phase behavior of the system towards higher values of $\phi$.}

\section{Summary and Conclusions} \label{s:conclusions}

In this article, we have studied the collective behavior of persistent active suspensions confined in asymmetric channels. The top boundary of the channel is a flat wall and its bottom boundary is composed of an array of funnel-like obstacles (see Fig. \ref{fig:scheme}). The active particles follow translational overdamped Langevin's dynamics and rotational run-and-tumble dynamics. 
\textcolor{black}{Our numerical results show}
a new type of collective motion in the shape of a wide and dense traveling band induced by the asymmetry of the boundaries of the confining channel. Such traveling band has not been observed neither in bulk nor under confinement in channels bounded by two flat walls.

By characterizing the mechanism of the traveling band's motion qualitatively and quantitatively, we identify its similarity with the motion of the tracks of vehicles, reason why we name this collective motion \textit{confinement-induced tracked locomotion}. While the band is moving, its internal structure can be described by 4 domains (see Fig. \ref{fig:P_theta}c). The mechanism relies on two key aspects: 1.) Particles in domain \textbf{II} constitute the thrust of the band by pushing domain \textbf{I}. 2.) Particles in domain \textbf{III} counter-flow in the lower part of the band via source-sink and vacancy diffusion mechanisms mediated by sliding along the obstacles. The existence of this counter-flow of particles facilitates the band's motion. Without it, the traveling band ends up in a clogging state.

Interestingly, we found the existence of a narrow range of obstacles tilt angles, $\theta_{obs}$ which promotes the emergence of the traveling state. We  discuss  the  preferred $\theta_{obs}$ in terms of an enhanced counter-current, $\Delta_x(y)$, as a consequence of an improved sliding along the obstacles for such angles. \horacio{We have discussed the effects of obstacles with a well-defined period on the top wall, by including an angle $\beta$ that gradually converts the top wall from flat to a trapezoidal wave. We observed that only small $\beta$ allows for the formation of the traveling structures, since the sliding of domains \textbf{I} and \textbf{II} along the top wall is crucial to the development of traveling states and high $\beta$ angles hinder the motility of active particles by promoting their accumulation inside the trapezoids, and effectively increasing the friction coefficient when particles slide along the top wall. Finally, we addressed the effects of increasing the channels' width, $L_y$, finding that in wider channels the active particles' packing fraction must be increased to recover the formation of traveling bands.}

We suggest that a possible experimental realization of the system might consist in a concentrated suspension of the motile microalgae \textit{Chlamydomonas reinhardtii} confined into a ring-shaped microfluidic device with one asymmetric boundary (with the shape of corrugations or funnels). It has been shown that a single cell  of \textit{C. reinhardtii} persistently follows  the curvature of a circular chamber\cite{bentley2022phenotyping}, which suggests that a collection of cells might behave in a similar way as described in the article. The ring-shaped chamber should have a perimeter of the same order of magnitude of the persistence length of the microalgae. To achieve high persistence lengths  and concentrate the cells in specific regions within the ring, \textit{C. reinhardtii} phototaxis can be leveraged. 
Another possible experimental realization of the proposed mechanism consists in Janus particles confined into rings. A recent study has reported clustering of Janus particles in ring-shaped soft confinement \cite{knippenberg2024motility}. Such clustering effect might be the basis for a traveling structure when the width of the ring and the packing fraction of the active suspension are correctly tuned. \horacio{The \textit{confinement-induced tracked locomotion} described in this article is persistent in long enough timescales (\textbf{see Figure S4}) to be appealing for technology.} A potential technological application could be the fast steered one-dimensional transport of passive objects along narrow channels. \horacio{In the coming future, we plan to address a detailed characterization of the transition to clogging states since it is relevant in granular flows in confined environments \cite{arevalo2016clogging,sinha2025facilitating,fang2024clogging}, and might be of interest to understand biofilms formation under confinement \cite{kumar2013microscale,conrad2018confined,fortune2022biofilm}. Another interesting study will be inducing preferential directions on the \textit{confinement-induced tracked locomotion} by adjusting the shape of the obstacles.}

\section{Supplementary Information}\label{s:SI}

\begin{itemize}
    \item \textbf{SI} containing additional figures and discussions. \textbf{Figure S1:} bulk, channels with 2 flat walls and channels with corrugations with no gaps. \textbf{Fig. S2:} pair correlation functions to choose the correct cut-off radius for the percolation criterion. \textbf{Figure S3:} Finite size effects, proof that in long channels the traveling band is still formed. \horacio{\textbf{Figure S4:} results of the time evolution of the center of mass of the system for a long simulation, showing inversion events and transient clogging states}  
    \item \textbf{Supplementary Video 1:} containing a video version of Figs. \ref{fig:diagrams} c,d and e.
    \item \textbf{Supplementary Video 2:} movie from simulations of the active suspension in bulk, confined within flat walls, and within channels with corrugations with no gaps.
    \item \textbf{Supplementary Video 3:} a movie showing the microscopic details of the motion of the traveling band.
    \item \horacio{\textbf{Supplementary Video 4:} a movie showing the active suspension confined into asymmetric channels as those presented in Fig. \ref{fig:Vcm-thetaObs}b: trapezoids with a characteristic angle $\beta$ in the top wall and funnel-like obstacles with an angle $\theta_{obs}$ in the bottom wall. $\theta_{obs} = 30^o$ and $\beta= 5^o$ (top), $30^o$ (central) and $60^o$ (bottom).}
    \item \horacio{\textbf{Supplementary Video 5:} a movie showing the active suspension confined into microchannels in which the period of the obstacles in the bottom wall, $l_{obs}$, is increased. $l_{obs}^0 =2\left( L_{obs}\cos\theta_{obs} + d + \sigma_{22}\right)$ with $\theta_{obs}= 30^o$, $L_{obs = 3\sigma_{22}}$ and $d = 1.25\sigma_{22}$. In the movie, $l_{obs} = 2l_{obs}^0$ (top), $l_{obs} = 3l_{obs}^0$ (central) and $l_{obs} = 4l_{obs}^0$ (bottom)}
    \item \textbf{Supplementary Data:} LAMMPS configurations of a system exhibiting \textit{confinement-induced tracked locomotion} in a long channel with $L_x = 10l_p$.

\end{itemize}

\begin{acknowledgments}

C.V. acknowledges fundings IHRC22/00002 and Proyecto PID2022-140407NB-C21 funded by
MCIN/AEI /10.13039/501100011033 and FEDER, UE. C.V. and H.S. acknowledge fundings from the European Union’s Horizon research and innovation programme under the Marie Skłodowska-Curie grant agreement No 101108868 (BIOMICAR). H.S and J. M-R. thanks the insightful discussions with Rodrigo Fern\'andez-Quevedo Garc\'ia. H. S. and C. V. acknowledge the discussions with Sujeet Kumar Choudhary and Marco Polin. \horacio{The authors acknowledge  insightful suggestions by Hugues Chat\'e.}
\end{acknowledgments}

\section*{Availability Statement}

The data that support the findings of this study are available from the corresponding author upon reasonable request.

\bibliography{apssamp}

\end{document}


\maketitle

\section{ The traveling band is only observed under confinement in asymmetric channels}

\begin{figure}[h!]
    \centering
    \includegraphics[width=1.0\linewidth]{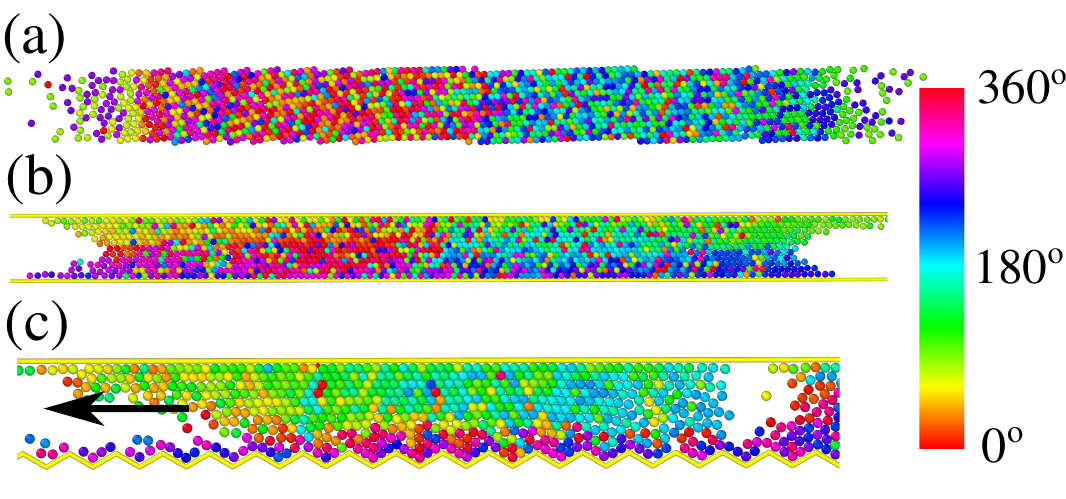}
    \caption{Configurations of active particles at the steady state with $\phi = 0.30$, $l_p = 10^5$. \textbf{(a)} Bulk: the system exhibits MIPS, and the band does not move persistently; \textbf{b} Confinement in a channel with flat walls: The system always evolves towards a clogging state; \textbf{(c)} Confinement in a channel with asymmetric boundaries (flat wall + corrugations with $\theta_{obs} = 30^\circ$): the system self-organizes into a traveling band whose mechanism of motion is similar to the cases discussed in the main text, \textit{confinement-induced tracked locomotion}. This Figure is complemented with the \textbf{Supplementary Video 1}.}
    \label{fig:S1}
\end{figure}

\clearpage

\section{Choosing $r_c$ for the percolation criterion}
\begin{figure}[h!]
    \centering
    \includegraphics[width=0.5\linewidth]{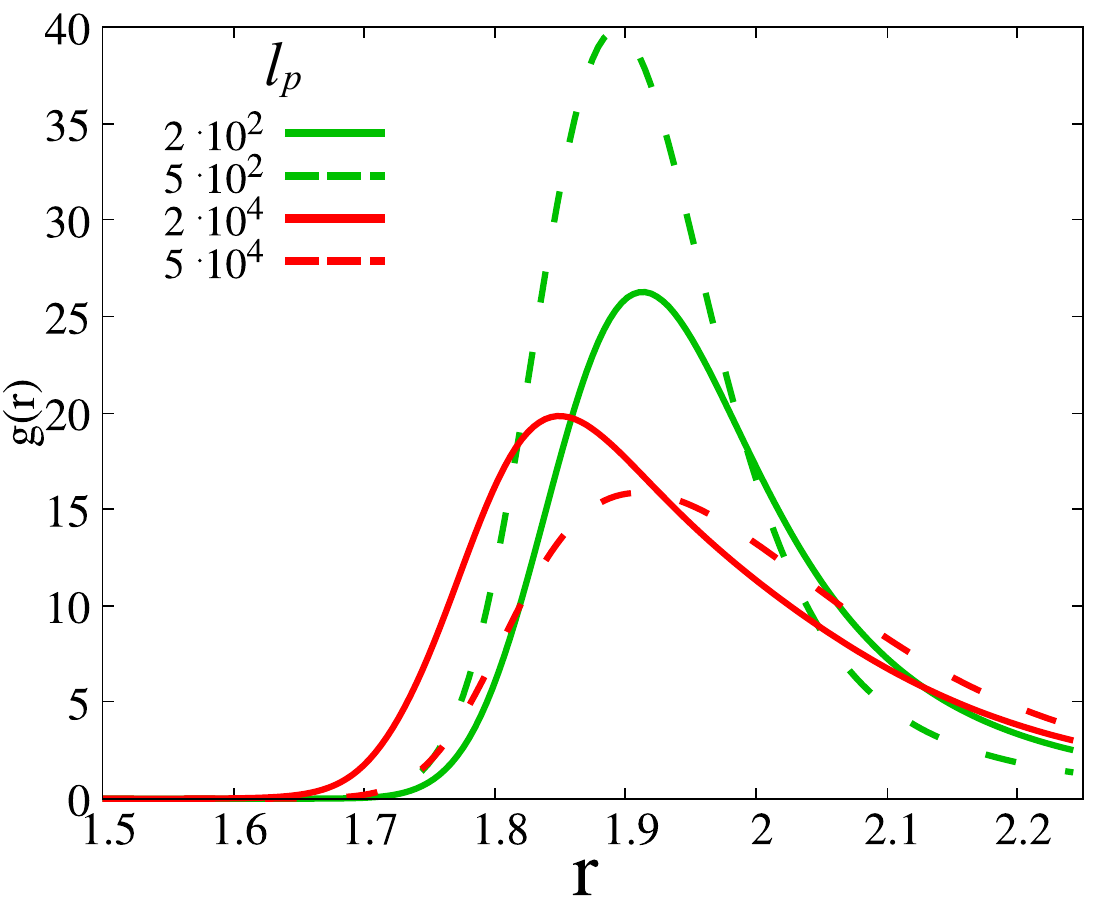}
    \caption{Radial distribution functions, $g(r)$, for $\phi=0.325$ and different persistent lengths for which either a clogging state (green) or a traveling state (red) is observed. With the aim of only detecting dense structures with the percolation criterion described in the main text, we choose $r_c = \sigma_{11} = 2\sigma_{22}$. This is the nominal contact distance between active particles, but greater than the real effective contact distance due to overlapping effects.}
    \label{fig:placeholder}
\end{figure}

\clearpage

\section{Finite Size Effects}

\begin{figure}[h!]
    \centering
    \includegraphics[width=1.0\linewidth]{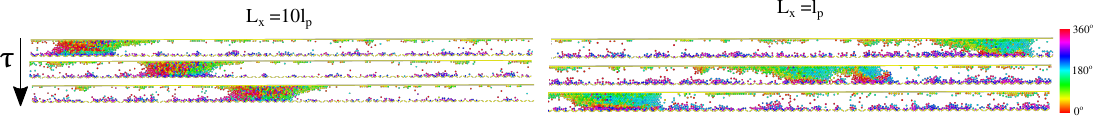}
    \caption{\textit{Confinement-induced tracked locomotion} is still observed in long channels, with length of the same order of magnitude (right panel) of the persistence length of the active particles, even one order of magnitude higher (left panel). The motion's mechanism  is exactly the same as described in the main text. However, in long channels, dense structures (traveling or not) may appear simultaneously at different positions. Thus there is a higher chance of encounters between two traveling structures moving in opposite directions, resulting in clogs, or traveling structures merging with clogs. For such reasons, the traveling structures survive for shorter time scales in longer channels. We add as \textbf{Supplementary Data} some LAMMPS configurations in which the traveling structure can be observed in a channel of $L_x = 10l_p$, initially formed at $x \approx 4200$, then it moves to the right until it merged with a clog. They can be visualized using OVITO or any other standard program for visualizing molecular simulation trajectories. As discussed in the main text, an experimental approach to realize short channels with PBC, is by confining the active suspension in ring-shaped chambers with perimeters of the order of magnitude of the persistence length of motile microorganisms. To achieve high persistence lengths, the photo- or chemotaxis of the microorganisms can be leveraged.} 
    \label{fig:S3}
\end{figure}

\clearpage

\section{Long simulations}

\begin{figure}[h!]
    \centering
    \includegraphics[width=1.0\linewidth]{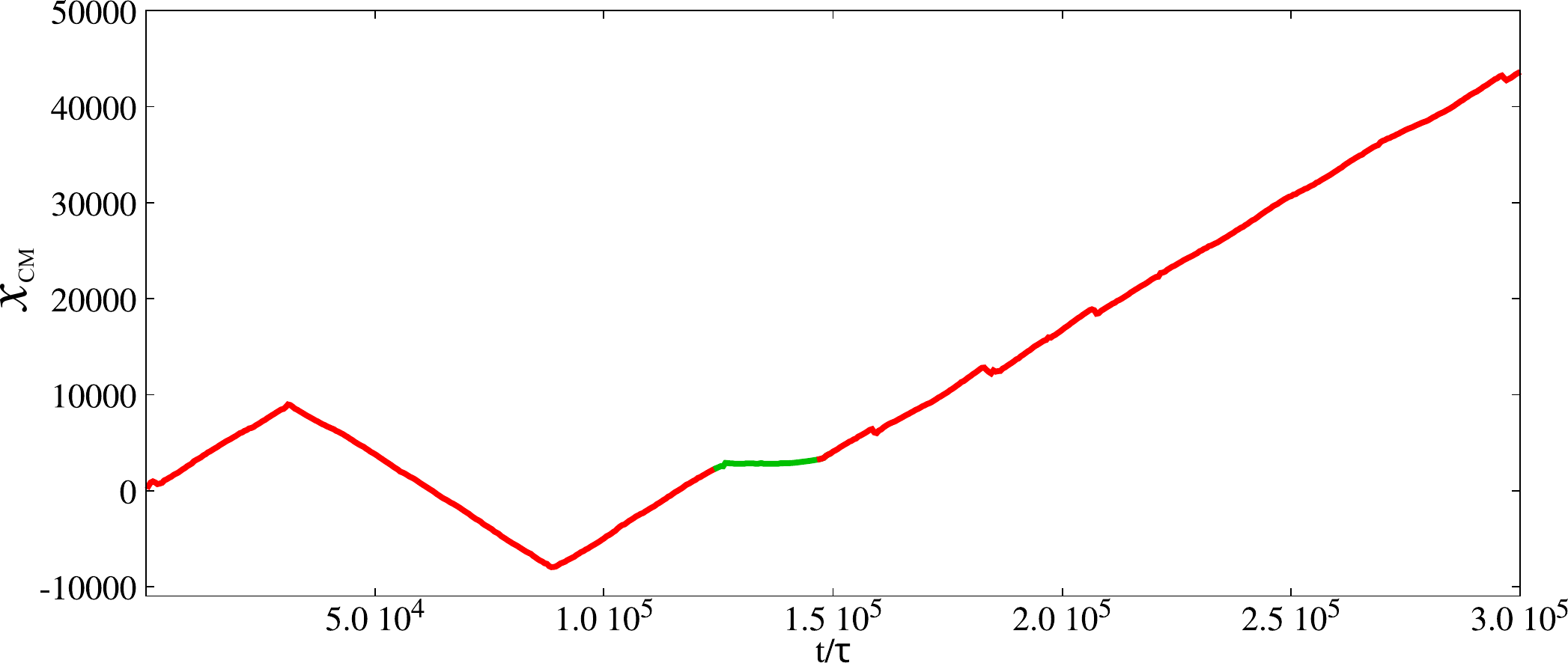}
    \caption{Time evolution of x component of the position of the center of mass of the system, $x_{CM}$ in a long simulation at $\phi = 0.30$, $l_p = 10^4$, $L_y = 10\sigma_{11}$ and $L_x \approx 800\sigma_{22}$. We can observe two inversion events at the beginning of the simulation and one clogging event that lasts about $30 000\tau$. The traveling band moves during $90\%$ of the simulation time, performing a quite persistent run of about $285 000\tau$ in the last part of the simulation. The run and tumble dynamics assigns a random angle in every tumble event, so the orientation distribution remains uniform throughout the simulation. After a clog is formed, it is only a matter of time for it to disassemble since the particles conforming the clog continue tumbling. Thus, in principle, in stability conditions of \textit{confinement-induced tracked locomotion}, during an infinite simulation, the system will evolve between three states: traveling band moving to the left, traveling band moving to the right, and a clog. Our results show that traveling states occur the vast majority of time in long simulations.} 
    \label{fig:S4}
\end{figure}